\documentstyle[12pt,aasms4]{article}

\begin{document}
\normalsize

\title{The Height Structure of the Solar Atmosphere from the EUV Perspective}

\author{J. Zhang, S. M. White, M. R. Kundu}

\affil{Dept. of Astronomy, University of Maryland, College Park, MD 20742}

\begin{abstract}

We investigate the structure of the solar chromosphere and transition 
region using full Sun images obtained with the Extreme Ultraviolet
Imaging Telescope (EIT) aboard the Solar and Heliospheric Observatory 
(SOHO) spacecraft. 
The limb seen in the EIT coronal images (taken in 
lines of Fe IX/X at 171 \AA, Fe XII at 195 \AA\
and Fe XV at 284 \AA) is an absorption 
limb predicted by models to occur at the top of the chromosphere 
where the density of neutral hydrogen becomes significant ($\sim10^{10}$ cm$^{-3}$).
The transition-region limb seen in He II 304 \AA\ images is an emission limb. 
We find: (1) the limb is higher at the poles than at the equator both in the
coronal images (by 1300 $\pm$ 650 km) and the 304 \AA\ images (by 3500 $\pm$ 1200 km); 
and (2) the 304 \AA\  limb is significantly higher than the limb in the
coronal images. The height difference is 3100 $\pm$ 1200 km at the equator,
and 6600 $\pm$ 1200 km at the poles. 
We suggest that the elevation of the 304 \AA\ limb above the limb in the coronal
images may be due to the upper surface of the chromosphere being bumpy,
possibly because of the presence of spicules.
The polar extension is consistent with a reduced heat input to the chromosphere
in the polar coronal holes compared with the quiet--Sun atmosphere at the
equator.

\end{abstract}


\clearpage

\section{Introduction}

The height structure of the solar atmosphere is an important feature of 
any model of the atmosphere and its energy balance. The models for the lower
atmosphere are largely based on the hydrostatic assumption and the idea of 
energy balance between energy loss, in the form of an outgoing radiation flux, 
and energy  input, possibly in the form of non-thermal energy generated by turbulent 
sub-photospheric motions for the chromosphere and
heat conduction from the hot corona for the transition region. 
The density and temperature gradients in the models are adjusted so that the predicted
radiation in various spectral ranges matches the observations (e.g.,
\cite{VAL81,FAL93,GJL97}).
The actual height structure in the chromosphere and transition region 
is difficult to resolve observationally, with the chromosphere being only of
order 3\arcsec\ thick in these models and the transition region as little as 0\farcs2 
thick.
Both may be obscured at the limb by spicules and by magnetic loops projecting into the corona.

The observational evidence that does exist for the height structure is largely 
inconsistent with the models. Most models put the
top of the chromosphere at heights ranging from 1700 to 2300 km above the
photosphere (e.g., \cite{AnA89,FAL93}), whereas
most observations in visible light during eclipses 
place it at about 5000 km (\cite{Zir96}).
Radio observations during eclipses may also be used to determine the height of
the chromosphere since short radio wavelengths become optically thick in the
chromosphere: these have given heights of 3400 km at $\lambda\,=\,$0.85 mm 
(\cite{EZJ93}), 6000 km at 1.3 mm (\cite{HHZ81}), 5500 km (\cite{BHG92})
and 8000 km (\cite{WhK94b}) at 3 mm.
Recently Johannesson and Zirin (1996) \nocite{JoZ96} measured the height of the 
chromosphere
based on disk images of high-pass H$\alpha$ filtergrams, and found
that the chromospheric height varies with heliographic latitude, being 4400 km
at the equator but about 6000 km at the poles.
Usually, the large observed extension (5000 km) has been ignored on the
grounds that it could be attributed to the numerous
spicules which clearly show up in off-band  H$\alpha$ images. 
However, Zirin (1996) has argued that the height is indeed 5000 km and that 
most of the theoretical models are likely 
to be wrong because they are based on the possibly incorrect 
assumption of hydrostatic equilibrium.
 
In this paper, we present measurements of the height of chromospheric features as 
well as of the transition region using full Sun images obtained with 
the Extreme Ultraviolet Imaging Telescope (EIT; \cite{DAB95,MCD97})
on the Solar and Heliospherical Observatory (SOHO) spacecraft.
Although the resolution of these images (2\farcs6 pixels) might not seem to lend itself
to such a study, the radius of the limb present in EIT images can repeatably be
measured with subpixel precision, and the height of the limb proves to be 
significantly above the photosphere.
The advantages of EIT observations for this purpose are twofold: (1) full disk
images provide us a chance to fit the entire limb of the lower solar atmosphere.
(2) EIT images are sensitive to transition region and coronal temperatures,
rather than chromospheric temperatures as in most other studies of the height of the
limb. 
We find, in stark contrast to the solar photosphere which is highly circular
(e.g., \cite{KBS98}),
that the heights of the polar and equatorial limbs are very different in both
the coronal and the transition region EIT images.
(While revising this paper we became aware of a similar result obtained
by Auch{\`e}re et al. 1998 using a different technique.)
\nocite{ADK98}
The sense of this latitude dependence is in agreement with the optical result
of Johanneson \& Zirin (1996)\nocite{JoZ96}.
In section 2 we discuss the nature of the limbs in the EIT images. 
In section 3, we present measurements and the results. The
final section discusses the implications of the results.
      
\section{The Solar Limb at EUV Wavelengths}

EIT obtains images in four
wavelength ranges centered on lines of the coronal species Fe IX/X (171 \AA), Fe
XII (195 \AA), and Fe XV (284 \AA), whose temperatures of maximum fractional abundance 
are about 1.0, 1.4 and 2.1 million K, respectively, 
and the transition--region diagnostic He II (304 \AA), whose temperature of maximum fractional
abundance is about 80,000 K. 
Inspection of EIT images indicates that the ``limb'' present in the
images made in the coronal lines is an absorbing or occulting limb, whereas
the limb in the 304 \AA\ image is an emission limb. This is to be expected,
since the 304 \AA\ image should be dominated by transition region gas which is
expected to be confined to a thin layer just above the chromosphere: thus 304
\AA\ should show a sharp fall--off in intensity above the chromosphere which
produces a well--defined emission limb.
The coronal lines, however, are optically thin and arise in a range of heights
throughout the corona: the limb in the coronal images 
appears where the lower atmosphere ceases to be
optically thick to EUV radiation and the intensity jumps by a factor of 2 as emission
from gas beyond the limb is added to that from gas in front of the limb.

Ionization of neutral H and He is the dominant absorption process for EUV photons
(e.g., \cite{CPB74}). At 200 \AA\ the cross--section for absorption is of
order 2 $\times$ 10$^{-19}$ cm$^2$ per neutral H atom (\cite{RBV94}). In an
exponential spherically symmetric atmosphere the line--of--sight column is
$n_{H0}\,\sqrt{2\pi\,h_0\,R_\odot}$, where $n_{H0}$ is the density of neutral
hydrogen at the lowest point in the solar atmosphere along the line of sight
and $h_0$ ($\ll\,R_\odot$) 
is the scale height of neutral hydrogen there. With a scale height of $h_0\ \sim$ 100 km,
which is plausible for the chromosphere, a neutral hydrogen density of only
10$^{10}$ cm$^3$ is required to make the atmosphere opaque. This number is
relatively insensitive to the actual scale height because of the square--root
dependence on $h_0$. In both standard hydrostatic--equilibrium ``VAL''
atmospheric models (e.g., \cite{VAL81,FAL93}) and empirically--determined
models (e.g., \cite{EZJ93,Zir96}) such a density of neutral H is reached very close to the top of 
the chromosphere, which implies that the absorption limb seen in the
EIT 171, 195 and 284 \AA\ images should be the top of the chromosphere.
The actual height in relation to the photosphere differs
greatly between the two models: it is 1500 -- 2000 km in the VAL models but
4700 km in Zirin's model. Note that the absorption cross--section at 284 \AA\ is larger at about 4 $\times$ 10$^{-19}$
cm$^2$, but given the small scale heights at the top of the chromosphere 
this is not sufficiently different that we expect to be able to
measure a limb height at 284 \AA\ which differs from the 171 and 195 \AA\
heights.

\section{Measurements and Results}

We analyze data sets composed of full--disk, full--resolution EIT synoptic
images in each of the four bandpasses acquired within a short time range 
(about 20 minutes) which have been taken several times daily 
since December 1996.
That the images are taken nearly simultaneously 
minimizes the measurement uncertainties due to variations in the instrument and solar features.
The measurements in this paper are based on 12 randomly--selected data sets evenly distributed between
April 1997 and January 1998.

The size of the limb is determined by fitting the limb to an ellipse.
Fitting is carried out by manually selecting
limb points, typically 40 per image (examples are shown in Figure 1).
Automated fitting routines do not generally work with high accuracy on EIT images
due to the irregularity of the limb: we found the manual method to be 
tedious but reliable.
The limb in these images is quite sharp:
the brightness jumps from the disk value to the off--limb value
within about three pixels along the radial direction.
Thus, it is not difficult to select good limb points manually. 
We avoided emission enhancements above the limb due to features such as 
loop or arcade structures in coronal images and prominences and 
macro-spicules in the 304 \AA\ images.
The coordinates of individual limb points are only accurate to within one pixel,
but the fitting procedure can determine the best--fit limb with sub--pixel accuracy.
By using test subjects other than the authors, we verified that this method
was repeatable and consistent. To improve statistical significance, the
fit was carried out three times for each image.
We determined the significance of ellipse
fits by comparing them with circle fits: e.g., on 1997 Aug 03 the
standard deviation of the distance of the selected points from the
best-fit ellipse at 304 \AA\ was 0.56 pixels, whereas for a circle it
was 0.93 pixels. The difference was less significant at coronal
wavelengths: an average of 0.43 pixels for the ellipse fit and 0.48 pixels
for the circle fit.

As expected, the fits to the absorption limb in the three coronal images were generally consistent with
one another: e.g., on 1997 Aug 03 the radius of the limb at the equator
was 365.7, 365.8, and 365.7 pixels in the 171 \AA, 195 \AA\ and 284 \AA\ 
images, respectively, while
in the polar direction the radius was 366.3, 366.4 and 366.6
pixels, respectively.
The radius of the limb in the coronal images 
is generally larger at the poles than at the
equator by a small amount: e.g., 0.70 pixels larger on 1997 Aug. 03. 

The emission limb in the  304 \AA\ image is always found to be larger 
than the limb in the coronal images, e.g. on 1997 Aug. 03 the 
limb at 304 \AA\ was 367.8 pixels at the equator and 369.8 pixels at the poles, which 
is 2.1 and 3.4 pixels larger than the corresponding limbs in the coronal
images.
In Figure 1 we compare the limbs in the 304 \AA\ and 195 \AA\ images directly for 1997 Aug.
03.  The limbs
at the poles and the east and west limbs are shown for both wavelengths,
together with the line which represents the best fit to the 195 \AA\ equatorial
limb.  The fact that the 195 \AA\ limb is well inside the 304 \AA\ 
transition-region-limb is clearly
evident, as is the fact that the polar limbs at 304 \AA\ are higher than the
equatorial limbs. The difference between the polar and equatorial limbs
in the coronal images is less than a pixel and is not readily apparent
in Fig.~1.

In order to determine the significance of these pixel differences,
we must first determine whether there is any 
asymmetry in the size of CCD pixels between the X (CCD rows) and Y (CCD columns)
directions.  For this purpose we analyzed a set of images taken 
on September 3, 1997, when the SOHO spacecraft was rotated
$90^\circ$ from its usual orientation.
(SOHO operations generally maintain  
an alignment accuracy of better than  $1^\circ$ between the CCD Y axis and
nominal solar rotation axis.)
A synoptic set of EIT observations were taken at about 07:00 UT in nominal orientation
(CCD Y-direction parallel to the solar rotation axis), and another at 13:00 UT
after the $90^\circ$ rotation (CCD Y-direction perpendicular to the 
solar rotation axis).
We fitted the limb size in both data sets. In the coronal images the average 
limb radii at the equator and poles, respectively, were
367.7 $\pm$ 0.4 and 368.4 $\pm$ 0.4 pixels before rotation and 
367.8 $\pm$ 0.4 and 368.3 $\pm$ 0.4 pixels
after rotation. The 304 \AA\  limb radii at the equator and poles,
respectively, were
369.3 $\pm$ 0.8 and 372.6 $\pm$ 0.8 pixels before rotation and 
369.3 $\pm$ 0.7 and 372.2 $\pm$ 0.7 pixels after rotation.
These data clearly indicate that to within our measurement accuracy the pixels
have identical X and Y dimensions on the sky, and that therefore the apparent
differences between the polar and equatorial dimensions cannot be attributed to
the pixel scale.
Thus, the latitude dependence of the limb heights must be real. 

We repeated the procedure described above for image sets on 11 other days 
almost evenly distributed over 7 months from April 1997 to January 1998. 
The relative height difference between the polar and equatorial limbs 
in the coronal images 
as a function of time is shown in Figure 2. It is significant and positive in
all data sets, ranging from 800 km to 1790 km with an average of 1300 km 
(with a measurement uncertainty of 650 km). 
The height of the limb in the 304 \AA\ images above the height of the limb 
in the coronal images as a function of  time 
is also shown in Figure 2. The range of height at the poles
is from about 5900 km up to 8100 km, while that at the equator is from
about 2100 km to 4100 km. Both are consistent with being constant in time to
within the measurement uncertainty (1200 km). The average 
height is 6600 $\pm$ 1200 km at pole and 3100 $\pm$ 1200 km at equator. 
The average difference between the polar and equatorial heights in the 304 \AA\
images is 3500 $\pm$ 1200 km.

The size of the coronal limb in pixels scales very well with the distance of
the SOHO spacecraft from the Sun, which varies by 3\% over the period shown
as SOHO orbits the Lagrangian L1 point between the Earth and the Sun.
Averaged over the 12 measurements, we find that one EIT pixel is 2\farcs610
$\pm$ 0\farcs002 in both dimensions if
the absorption limb in the coronal images is the photosphere (6.96 $\times$
10$^5$ km). Note that the statistical uncertainty in this measurement is
smaller than 0.1\%, indicating that the feature responsible for the coronal
limb is very stable over the 7 months. If the absorption is in
the chromosphere at a height of 2000 km, the pixel scale would be 2\farcs617.

\section{Discussion and Conclusion}

We have fit the solar limb in sets of EIT  171, 195, 284 and 304 \AA\ synoptic
images and find that (i) the three coronal lines images, where the limb
is seen in absorption, yield an identical elliptic disk (to within our
uncertainty) with the polar limb being 1300 km larger than the equatorial limb
on average (ii) the 304 \AA\ images, in which the transition-region limb is 
seen as an emission feature,
invariably show a disk larger than in the coronal images, and also larger at
the poles than at the equator.
By detailed inspection of limb profiles of individual cases, we have determined that the 
radius we measure for the 304 \AA\ limb
corresponds to a point on the upper part of the decrease in intensity at
the limb, whereas in the coronal images the radius we measure
corresponds to a point near the base of the limb--brightening feature,
i.e., just above the point where the intensity starts to increase sharply
with radius at the limb.
A careful inspection of average profiles at the limbs indicates that the peak in,
e.g., the 195 \AA\ images occurs at a slightly greater height than the peak in the 304
\AA\ images, but that the rise towards the peak in the 195 \AA\ images occurs
well below the start of the fall--off in the 304 \AA\ images. 
However, the average limb profiles are
clearly not consistent with both the 304 \AA\ and 195 \AA\ images having a sharp
edge at about the same height: there must exist 304 \AA--emitting
material above the coronal boundary. Note that the 304 \AA\ limb
determination is not affected by the contribution of coronal lines in
the 304 \AA\ bandpass: the relative contribution of the coronal lines
can easily be determined by comparison with coronal images well above
the limb, and is not significant at the limb.

A polar extension of the chromosphere would be consistent with a reduced heat
input to the coronal hole atmosphere compared with the quiet Sun.
Hydrostatic models for the solar atmosphere predict a more extended
chromosphere when the energy supplied to the chromosphere is reduced, 
e.g. the height of VAL model A (dark
point within a cell) is 250 km larger than that of VAL model F 
(very bright network element) and 500 km larger than that of VAL model P (plage area
with medium brightness). The model of Basri et al (1979) \nocite{BLB79} predicts
that the height of the chromosphere in a dark point is 1000 km larger than that 
of average quiet Sun.
This occurs because the atmosphere must relax to a state in which
the energy loss matches the energy input: since radiative energy loss varies
as $n^2$, a more extended chromosphere produces smaller $n$ for the same mass
column and therefore reduces radiative losses.
In this sense a larger chromospheric height at the poles would be consistent
with the expected small heat input from the polar coronal holes

The 304 \AA\ emission  must arise in a region where the ambient He is singly
ionized, which corresponds to a temperature of order 80000 K.
If we assume that the transition region is truly as thin as all models
predict and is a uniform sphere, then the solar limb in the 304 \AA\ images 
should represent the top of the chromosphere. However, earlier we showed that
the limb in coronal images should also be
found at the top of the chromosphere where significant amounts of neutral H
are present to absorb the EUV lines. The difference in height between the 304
\AA\ limb and the limbs in the three coronal images is clearly not consistent with
a model in which the chromosphere is a well--defined layer of uniform thickness
and the 304 \AA\ emission arises in a thin spherical surface at its upper edge.

The fact that the coronal limb lies {\it well below} the He II 304 \AA\ limb
requires either that (i) there be material at coronal temperatures extending to
heights below the top of the surface on which 304 \AA\ emission is to be found, or
that (ii) all the coronal material lies radially above the 304 \AA\ limb but
the layer in the chromosphere which is
opaque in the EUV and thus defines the coronal limb lies well below the 304 \AA\
emitting layer, permitting a transparent window below the 304 \AA\ limb
through which coronal material beyond and in front of the limb may be seen.

The first of these two explanations suggests the plausible model of a bumpy
interface between the corona and the chromosphere: effectively the height of
the top of the chromosphere varies from place to place by up to several
thousand kilometers, with the 304 \AA\ emitting layer forming a skin (the
transition region) on the upper surface of the bumpy chromosphere. The limb at 304
\AA\ will then appear to be at the upper envelope of this surface, since 304
\AA\ is in emission, while the coronal limb will be at the bottom envelope of
the bumpy surface as long as the chromospheric fingers which protrude above
the lower envelope do not produce significant absorption of the EUV
radiation. The neutral hydrogen density in these ``fingers'' is likely to be
lower than in the bulk of the chromosphere since they are surrounding by
coronal material emitting ionizing radiation. The larger difference in height
between the 304 \AA\ limb and the coronal limb at the poles then implies that
the top of the chromosphere is much ``bumpier'' at the poles than at the equator.
The ``bumpiness'' of the top of the chromosphere could well be associated with spicules.

The second explanation requires that the level of ionization of H in the upper
chromosphere be much higher than found in any current models of the
chromosphere, since only if the density of neutral
H and He is low can the upper chromosphere
below the 304 \AA\ emitting layer be transparent to EUV radiation at 171 -- 284 \AA.
We note that He$^+$ can be ruled out as the main absorber since its absorption
edge occurs at 228 \AA\ and thus 284 \AA\ photons are not absorbed by He$^+$.

One important unanswered question is whether
the extension of the chromosphere at the pole is a local characteristic 
associated with only coronal holes, or a global feature which is associated
with the poles but not directly with the presence of coronal holes there.
The EIT images do not seem to be suitable to address this question: we would
like to look for an extension of the 304 \AA\ limb over a coronal hole at the
equator, but locally it is a much harder task to identify and fit the limb
than it is in the global sense which we have used here.
Johannesson and Zirin (1996) in their $H_{\alpha}$ observations found
that the polar height excess is co-spatial with the polar coronal holes,
and that a similar excess is also found within active regions outside the polar region,
although this may be unrelated.

\section{ACKNOWLEDGMENTS}
This research was supported by NASA grant NAG-5-6257 and NSF grant ATM-96-12738.
We thank Markus Aschwanden and Alexander Nindos for their help in testing the
fitting routine, and Ken Dere, Markus Aschwanden, Jeff Newmark, Bill
Thompson and James A. Klimchuck for valuable discussions.

\clearpage
\centerline{\bf FIGURE CAPTIONS}

\figcaption{Comparison of the limb location in the EIT 304 \AA\ (left panels) and
195 \AA\ (right panels) on 1997 August 3. Each panel is a 1040\arcsec\ $\times$
156\arcsec\ (400 $\times$ 60 pixels in the full--resolution images) 
region from either the north or south pole or the east or west
limb, as marked. The relevant portion of the image has been rotated so that
the limb is at the top of the image. In each panel,
the equatorial limb from the fit of the 195 \AA\ image
has been plotted as a line (solid white line in the 304 \AA\
panels, dotted black line in the 195 \AA\ images), while pixels used in
fitting the limbs are shown by crosses.
Note that the fields of view in the corresponding 304 \AA\ and 195 \AA\ panels
may differ by up to a pixel due to rounding.
\label{fig1}}

\figcaption{Heights of features in the EIT images: 
the height differences between the 304 \AA\ limb 
and the limb in the coronal images
at (i) the poles are marked by triangles, while the differences at (ii)
the equator are marked by stars. (iii) The height of the polar limb above
the equatorial limb in the coronal
images is marked by filled circles. Measurements are
plotted for each of the 12 data sets investigated.
Error bars are plotted at $\pm$ $\sigma$, where $\sigma$ is the uncertainty in
the fit of the limb points to an ellipse. 
\label{fig2}}

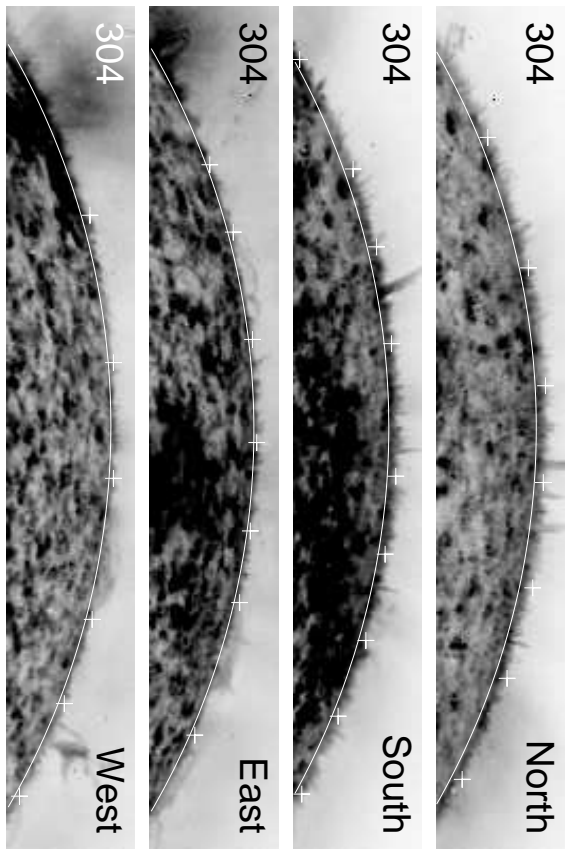
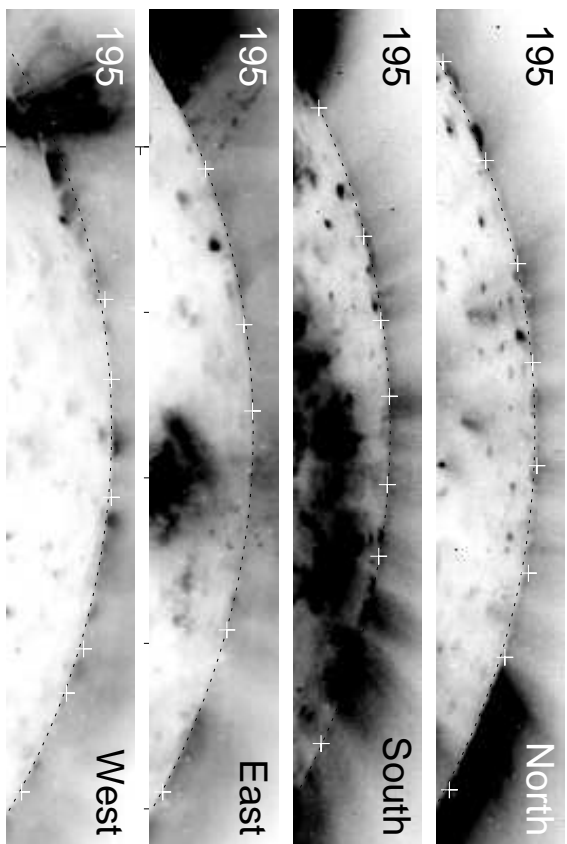
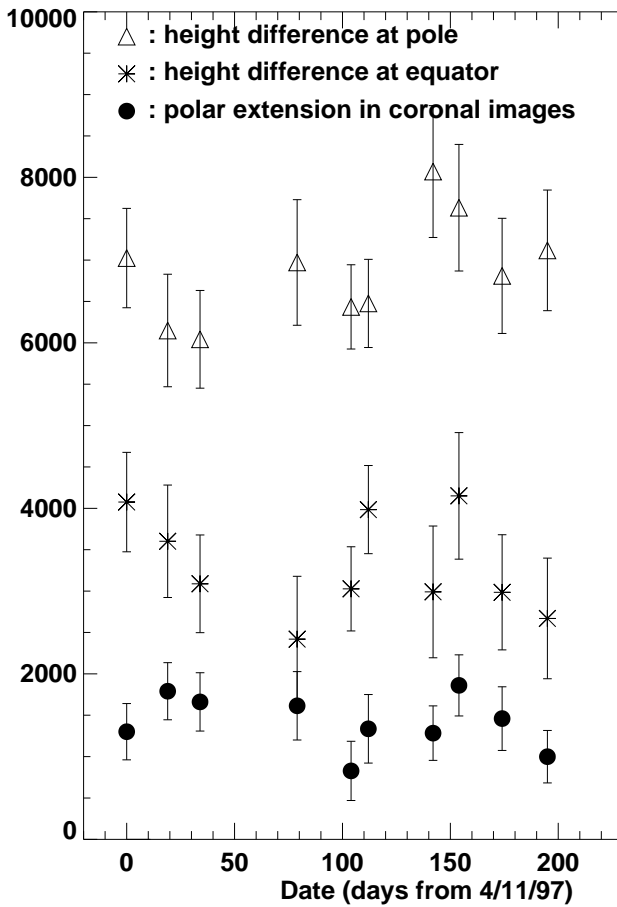
\end{document}